\RequirePackage{ifpdf}
\ifpdf % We are running pdfTeX in pdf mode
\documentclass[pdftex]{sigma}
\else
\documentclass{sigma}
\fi

\begin{document}

%\allowdisplaybreaks

\renewcommand{\PaperNumber}{088}

\FirstPageHeading

\ShortArticleName{Trigonometric Solutions of WDVV Equations and Generalized
CMS Systems}

\ArticleName{Trigonometric Solutions of WDVV Equations\\ and Generalized
Calogero--Moser--Sutherland  Systems}

\Author{Misha V. FEIGIN}

\AuthorNameForHeading{M.V. Feigin}

\Address{Department of Mathematics, University of Glasgow, G12 8QW, UK}
\Email{\href{mailto:M.Feigin@maths.gla.ac.uk}{M.Feigin@maths.gla.ac.uk}}
\URLaddress{\url{http://www.maths.gla.ac.uk/~mf/}}

\ArticleDates{Received May 18, 2009, in f\/inal form September 07, 2009;  Published online September 17, 2009}

\Abstract{We consider trigonometric solutions of WDVV equations and derive
geometric conditions when a collection of vectors with
multiplicities determines such a solution. We incorporate these
conditions into the notion of trigonometric Veselov system
($\vee$-system) and we determine all trigonometric $\vee$-systems with
up to f\/ive vectors. We show that genera\-li\-zed
Calogero--Moser--Sutherland operator admits a factorized eigenfunction
if and only if it corresponds to the trigonometric $\vee$-system; this inverts a one-way implication observed by Veselov for the rational solutions.
}


\Classification{35Q40; 52C99}

\newcommand{\p}{\partial}

\renewcommand{\a}{\alpha}
\renewcommand{\b}{\beta}
\newcommand{\g}{\gamma}

\newcommand{\R}{\mathbb R}
\newcommand{\Q}{\mathbb Q}
\newcommand{\C}{\mathbb C}
\newcommand{\N}{\mathbb N}
\newcommand{\Z}{\mathbb Z}
\renewcommand{\t}{\widetilde}
\renewcommand{\hat}{\widehat}
\renewcommand{\tilde}{\widetilde}

\newcommand{\ol}[1]{\overline{#1}}
\newcommand{\cl}{{\cal L}}
\newcommand{\cR}{{\cal R}}
\newcommand{\cA}{{\cal A}}

\newcommand{\nad}[2]{\genfrac{}{}{0pt}{}{#1}{#2}}

\section{Introduction}

In this note we are interested in the special trilogarithmic
solutions of the generalized Witten--Dijkgraaf--Verlinde--Verlinde (WDVV)
equations \cite{MMM}. Such solutions are determined by a f\/inite
collection $A$ of covectors $\a$ with multiplicities $c_\a$. More
specif\/ically, the prepotential satisfying the WDVV equations has the
form \begin{gather}\label{fintro}
F= \sum_{\a \in A} c_\a Li_3\big(e^{2i\a(x)}\big) +
\mbox{cubic terms}, \end{gather}
where $Li_3$ is the trilogarithm function. Solution of this type for the
$A_n$ root system appeared in \cite{MMM2} in relation with
Seiberg--Witten theory. More systematically such solutions were
studied by Hoevenaars and Martini in \cite{H1, H2} who
determined solutions for all irreducible reduced root systems
\cite{H2}. More recently solutions of the form \eqref{fintro} were
derived from reductions of Egorov hydrodynamic chains in \cite{P}.

The rational versions of solutions \eqref{fintro} play an important
role in the theory of Frobenius manifolds, a geometric framework for the
WDVV equations~\cite{D}. Thus solutions corresponding to the Coxeter
root systems are almost dual to the Frobenius structures on the
orbit spaces of f\/inite Coxeter groups \cite{D2}. In the
trigonometric case such a duality is verif\/ied for the af\/f\/ine $A_n$
case in \cite{R,RS}. The study of general rational solutions
of the form
\begin{gather}\label{fintro2} F=\sum_{\a\in A} \a(x)^2 \log \a(x) \end{gather}
was initiated by Veselov in \cite{V1} where a geometric notion of
the $\vee$-system equivalent to the WDVV equations for
\eqref{fintro2} was introduced. It was shown in \cite{V3} that any
generalized Calogero--Moser--Sutherland (CMS) operator admitting a
factorized eigenfunction determines a
$\vee$-system.

In this note we are interested in the solutions \eqref{fintro} where
the cubic terms involve extra variable like in the works \cite{H1,H2} on the solutions for the root systems. We derive geometric
and algebraic conditions for a system of vectors with multiplicities
so that the corresponding function \eqref{fintro} satisf\/ies the WDVV
equations. These conditions should be thought of as trigonometric
analogue of the notion of the $\vee$-system. The conditions carry
rather strong geometrical restrictions on the collection of vectors
formulated in terms of series of vectors parallel to a chosen one.
We illustrate this by determining all trigonometric $\vee$-systems
with up to f\/ive vectors in the plane.

Trigonometric ansatz, in contrast to the rational one, allows to
def\/ine the generalized CMS operator corresponding to the solution~\eqref{fintro}. We show that this operator has a factorized
eigenfunction. This statement inverts the one for the rational
$\vee$-systems obtained in~\cite{V3}. In fact our arguments follow~\cite{V3} very closely. We also discuss additional condition needed
to have trigonometric solution to the WDVV equations starting from a CMS
operator with factorized eigenfunction.

\section[Trigonometric $\vee$-systems]{Trigonometric $\boldsymbol{\vee}$-systems}

Consider a function $F$ of the form
\begin{gather}\label{F}
F=\frac13 y^3 +
\sum_{\a\in A} c_{\a} \a(x)^2 y + \lambda \sum_{\a \in A} c_\a
f(\a(x)), \end{gather}
where $A$ is a f\/inite collection of covectors on
$V\cong\C^n$, $x=(x_1,\ldots,x_n)$, $c_\a$, $\lambda$ are non-zero
constants and function $f(x)$ satisf\/ies $f'''(x)=\cot x$. The last
equation f\/ixes function $f(x)$ up to 2nd order terms which will not
be important for the WDVV equations below. We may f\/ix a~choice of
$f(x)$ by
\[
f(x)=\frac16 i x^3 + \frac14 {\rm Li}_3\big(e^{-2ix}\big).
\]

 The ansatz \eqref{F}
 introducing extra variable $y$ was proposed in \cite{H2} in the
case of root systems $A=\mathcal R$. The form \eqref{F} guarantees
that the matrix of third derivatives involving $y$ is constant, as we
will explore below.

{\it We will assume throughout the paper that collection $A$ of
covectors $\a$ belongs to an  $n$-dimen\-sional lattice, and that the
bilinear form \begin{gather}\label{G} (u,v):=\sum\limits_{\a\in A}c_\a \a(u) \a(v) \end{gather} is
non-degenerate on~$V$.} The form $(\cdot,\cdot)$ identif\/ies $V$ and
$V^*$ and following \cite{V1} we will denote by~$\gamma^\vee$ the vector dual to the
covector $\gamma$. We will also denote through
$(\cdot,\cdot)$ the corresponding inner product on $V^*$. We are
interested in the conditions on $\{\a, c_{\a}, \lambda\}$ when
function $F$ satisf\/ies the WDVV
    equations \begin{gather}\label{wdvv} F_i F_k^{-1} F_j = F_j F_k^{-1} F_i, \end{gather}
$i,j,k=0, 1,\ldots,n$. Here $F_i$ are $(n+1)\times (n+1)$ matrices
of third derivatives in the coordinates $(x_0=y,x_1,\ldots,x_n)$,
$(F_i)_{ab}=\frac{\p^3F}{\p x_i \p x_a \p x_b}$. It is suf\/f\/icient to
f\/ix $k=0$, then $F_0=F_y$ is the following non-degenerate matrix
\[
F_y=2 \left(
\begin{array}{cc}
1 & 0\\
0& \sum\limits_{\a \in A} c_\a \a \otimes \a
\end{array}
 \right).
\]
Similarly
\[
F_i=\left(
\begin{array}{cc}
0 & 2\sum\limits_{\a \in A} c_\a \a_i \a\vspace{2mm}\\
 2\sum\limits_{\a \in A}  c_\a \a_i \a & \lambda\sum\limits_{\a \in A} c_\a \a_i \cot \a(x) \a \otimes \a
\end{array}
 \right),
\]
 where we denoted by $\a$ both column and row vectors
 $\a=(\a_1,\ldots,\a_n)$.

The WDVV conditions for a function $F$ can be reformulated partly
using geometry of the system $A$. For any $\a \in A$ let us collect
all the covectors from $A$ non-parallel to $\a$ into the disjoint union of {\it
\bf  $\pmb \a$-series} $\Gamma_\a^1, \ldots, \Gamma_\a^k$. These
series are determined by the property that for any $s=1,\ldots, k$
and for any two covectors $\gamma_1, \gamma_2 \in \Gamma_\a^s$ one
has either $\gamma_1-\gamma_2=n\alpha$ or
$\gamma_1+\gamma_2=n\alpha$ for some {\it integer} $n$. We also
assume that the series are maximal, that is if $\b \in \Gamma_\a^i$
then $\Gamma_\a^i$ must contain all the vectors of the form $\pm \b
+n \a \in A$ with $n \in \Z$.

We note that solution \eqref{F} is not af\/fected if some of the
covectors $\a \in A$ are replaced with~$-\a$. By appropriate choice
of signs the vectors can be made to belong to a half-space, we will
denote such systems as~$A_+$. Moreover, for any $\a\in A$ one can
choose a positive system $A_+ \ni \a$ in such a way that $\a$-series
$\Gamma_\a^s$ will consist of vectors of the form $\b_s + n_i \a \in
A_+$ for appropriate integer parameters $n_i$ with $\b_s \in A_+$.

\begin{definition} Let $A \subset V^*\cong\C^n$ be a f\/inite
collection of covectors $\a$ with multiplicities $c_\a$ such that
the corresponding form \eqref{G} is non-degenerate and the covectors
$\a$ belong to an $n$-dimensional lattice. We say that $A$ is a
trigonometric $\vee$-system if for any $\a \in A$ and for any
$\a$-series $\Gamma_\a^s$ one has \begin{gather}\label{V1} \sum_{\b \in \Gamma_\a^s}
c_\b (\a,\b) \a\wedge \b=0. \end{gather}
\end{definition}

Notice that $\a \wedge \beta_1 = \pm \a\wedge \b_2$ if $\b_1$, $\b_2$
belong to the same $\a$-series $\Gamma_\a^s$ so identities \eqref{V1}
may be simplif\/ied accordingly. Also replacement of some of the covectors
by their opposite preserves the class of trigonometric
$\vee$-systems. Note also that the non-degenerate linear transformations
act naturally on the trigonometric $\vee$-systems, and that the direct
sum $A_1 \oplus A_2$ of the trigonometric $\vee$-systems $A_1
\subset V_1^*$, $A_2 \subset V_2^*$ considered as a set of covectors
in $V_1\oplus V_2$ is again a trigonometric $\vee$-system. The
systems obtained in this way will be called {\it reducible}. If such
a decomposition is not possible then the (trigonometric
$\vee$-)system is called irreducible.

\begin{theorem}\label{t1}
The WDVV equations \begin{gather*}%\label{wdvv1}
F_i F_y^{-1} F_j = F_j F_y^{-1} F_i,
\end{gather*} $i,j=0, 1,\ldots,n$,
 for the function \eqref{F} are equivalent to the following two
 conditions:
 \begin{enumerate}\itemsep=0pt
\item[$1)$]  $A$ is a trigonometric
 $\vee$-system;
\item[$2)$] for a positive system $A_+$ and for any vectors $a,b,c,d \in V$
 \begin{gather}\label{V2} \sum_{\a,\b \in A_+}
\left(\frac14 \lambda^2   (\a, \b) - 1\right) c_\a c_\b B_{\a,\b}(a,b)
B_{\a,\b}(c,d) =0, \end{gather} where $B_{\a,\b}(a,b)=\a \wedge \b (a,b)=
\a(a)\b(b)-\a(b)\b(a)$.
\end{enumerate}
\end{theorem}

\begin{proof}
For a vector $a \in V$ we def\/ine $F_a^\vee = F_y^{-1}
F_a$ where $F_a= \sum\limits_{i=1}^n a_i F_i$. The WDVV equations are
equivalent to the commutativity $[F_a^\vee, F_b^\vee]=0$ for any
$a,b \in V$. We have
\[
F_i^\vee=\left(
\begin{array}{cc}
0 & \sum\limits_{\a \in A}  c_\a \a_i \a\vspace{2mm}\\
 \sum\limits_{\a \in A}  c_\a \a_i \a^\vee & \frac{\lambda}{2} \sum\limits_{\a \in A} c_\a \a_i \cot \a(x) \a \otimes
 \a^\vee
\end{array}
 \right),
\]
where $\a^\vee$ is the (column) vector dual to the (row) covector
$\a$ under the bilinear form $G=\sum\limits_{\a \in A} c_{\a} \a\otimes\a$.
Therefore
\[
F_a^\vee=\left(
\begin{array}{cc}
0 & \sum\limits_{\a \in A}  c_\a \a(a) \a \vspace{2mm}\\
 \sum\limits_{\a \in A}  c_\a \a(a) \a^\vee & \frac{\lambda}{2} \sum\limits_{\a \in A} c_\a \a(a) \cot \a(x) \a \otimes
 \a^\vee
\end{array}
 \right)
\]
for any $a \in \C^n$. Now the product $F_a^\vee F_b^\vee$ equals
\[
 \left(
\begin{array}{cc}
\sum\limits_{\a,\b \in A} c_\a c_\b \a(a) \b(b) \a(\b^\vee) &
\frac{\lambda}{2} \sum\limits_{\a,\b \in A}  c_\a c_\b \a(a) \b(b) \a(\b^\vee) \cot \b(x) \b \vspace{2mm}\\
\frac{\lambda}{2} \sum\limits_{\a, \b \in A}  c_\b c_\a \a(a) \b(b)
\a(\b^\vee) \cot \a(x) \a^\vee  & \nad{\sum\limits_{\a,\b \in A} c_\a c_\b
\a(a) \b(b) \a^\vee \otimes \b }{+\frac{\lambda^2}{4} \sum\limits_{\a, \b
\in A} c_\a c_\b \a(a) \b(b) \a(\b^\vee) \cot \a(x) \cot \b(x)
\b\otimes\a^\vee}
\end{array}
 \right).
\]
Therefore $[F_a^\vee, F_b^\vee]=0$ is equivalent to the identities
\begin{gather}\label{sing0}
\sum_{\a,\b \in A} c_\a c_\b B_{\a,\b}(a,b) (\a, \b)
\cot \a(x) \a^\vee = 0, \\
 \label{sing} \sum_{\a,\b \in A}  \left( \frac{\lambda^2}{4} c_\a
c_\b (\a, \b) \cot \a(x) \cot \b(x) + c_\a c_\b  \right)
B_{\a,\b}(a,b) \a\wedge \b=0.
\end{gather} To cancel singularities in
\eqref{sing} one should have
\[
\sum_{\nad{\b \in A}{\b \nsim \a}}  c_\b (\a, \b) \cot \b(x)
B_{\a,\b}(a,b) \a\wedge \b=0
\]
when $\cot \a(x)=0$. A linear combination of functions $\cot
\b(x)|_{\cot \a(x)=0}$ can vanish only if it vanishes for each
$\a$-series:
\[
\sum_{\b\in \Gamma_\a^s}  c_\b (\a, \b) \cot \b(x) B_{\a,\b}(a,b)
\a\wedge \b=0
\]
for all $\a$-series $\Gamma_\a^s$ (see e.g.~\cite{F} for more
detailed explanation). The last relation can be simplif\/ied~as
 \begin{gather}\label{V11} \sum_{\b
\in \Gamma_\a^s} c_\b (\a,\b) \a\wedge \b=0, \end{gather} which means that
$A$ is a trigonometric $\vee$-system.
 Identities \eqref{V11} guarantee that the
left-hand side of \eqref{sing} is non-singular. Since all the vectors
from $A$ belong to an $n$-dimensional lattice with basis $e^1,
\ldots, e^n$, the left-hand side of \eqref{sing} is a rational
function in the exponential coordinates~$e^{e^i(x)}$. This rational
function has degree zero and therefore it is a constant. We can
assume that all covectors from $A$ belong to a half-space hence form
a positive system $A_+$, so in appropriate limit $\cot(\a,x) \to i$
for all $\a\in A_+$. Thus property \eqref{sing} is equivalent to
\eqref{V11} together with the condition
 \begin{gather*} %\label{V12}
 \sum_{\a,\b \in A_+}
\left(\frac{\lambda^2}{4} c_\a c_\b  (\a, \b) - c_\a c_\b \right)
B_{\a,\b}(a,b) \a\wedge \b =0.
\end{gather*}
 The remaining condition
\eqref{sing0} is equivalent to the set of properties \begin{gather}\label{V3}
\sum_{\b \in A} c_\b (\a, \b) B_{\a,\b}(a,b) = 0, \end{gather} for any $\a
\in A$. Identities \eqref{V3} follow from the $\vee$-conditions
\eqref{V11}, this completes the proof of the theorem.
\end{proof}

\begin{remark}\label{remark1} Let trigonometric $\vee$-systems $A_1 \subset V_1^*,
A_2\subset V_2^*$ def\/ine the solutions \eqref{F} of the WDVV
equations for some $\lambda_1$, $\lambda_2$. Then the trigonometric
$\vee$-system $A_1 \oplus A_2$ does not def\/ine a~solution. Indeed,
let us take vectors $a,c \in V_1$ and $b,d \in V_2$. Then property
\eqref{V2} implies that \begin{gather}\label{dirsn} (a,c)_1 (b,d)_2 =0, \end{gather} where
$(\cdot,\cdot)_{1,2}$ are $\vee$-forms \eqref{G} in the corresponding
spaces $V_{1,2}$. Clearly, the relation \eqref{dirsn} does not hold
for general vectors $a$, $b$, $c$, $d$.
\end{remark}

\begin{remark}\label{remark2} Not all the trigonometric solutions of the WDVV
equations have the form~\eqref{F}. It is shown in \cite{BMMM} that
trilogarithmic functions have to arise when ansatz for $F$ is given
by summation of $g((\a,x))$ over the roots of a root system, $x\in
V$.
\end{remark}

\begin{remark}\label{remark3} A slightly more general ansatz for the solutions $F$ can be considered when cubic terms in $x$ are added to $F$. Similarly to the proof of Theorem~\ref{t1} it follows that $A$ still has to be a trigonometric $\vee$-system. The almost dual potentials corresponding to the $A_n$ af\/f\/ine Weyl group orbit spaces have such a form~\cite{RS}. The corresponding trigonometric $\vee$-system $A$ is the $A_n$ root system  in this case.
\end{remark}

\begin{proposition}\label{proposition1}
Let $A=\{\a, c_\a\}$ be a trigonometric $\vee$-system. Then the set
of vectors $\{\sqrt{c_{\a}}\a\}$ is a $($rational$)$ $\vee$-system, that
is $F^{\rm rat}=\sum\limits_{\a\in A} c_\a \a(x)^2 \log \a(x)$ is a solution of
the WDVV equations in the space $V$.
\end{proposition}

\begin{proof} By def\/inition of the trigonometric $\vee$-system for
any $\a \in A$
 relations \eqref{V1} hold. Consider two-dimensional plane $\pi \subset V$ and sum up relations \eqref{V1} over $s$ so that
the $\a$-series $\Gamma_\a^s$ belong to the plane $\pi \ni \a$. We
arrive at the relations
\[
 \sum_{\b \in A\cap\pi} c_\b (\a,\b)
\a\wedge \b=0,
\] or, equivalently, \begin{gather}\label{rV} \sum_{\b \in A\cap\pi}
c_\b (\a,\b)\b \quad \mbox{ is proportional to } \a. \end{gather} Relations
\eqref{rV} is a def\/inition of the (rational) $\vee$-system for the
set of covectors $\{\sqrt{c_{\a}}\a\}$ (see~\cite{V1} and~\cite{FV2} for the complex space). It is equivalent to the property
that $F^{\rm rat}$ satisf\/ies WDVV equations in the space $V$ \cite{V1,FV2}. Proposition is proven.
\end{proof}

Due to existence of extra variable $y$ in the ansatz \eqref{F} the
WDVV equations are nontrivial already when $n=2$. Thus it is natural
to study at f\/irst two-dimensional conf\/igurations $A$ def\/ining
solutions of WDVV equations. When $A$ consists of one vector the
corresponding form~\eqref{G} is degenerate. If $A$ consists of two
non-collinear vectors $\a$, $\b$ then it follows that $(\a,\b)=0$
therefore relation \eqref{V1} holds and $A$ is a trigonometric
$\vee$-system. However relation \eqref{V2} cannot hold then for any
$\lambda$ and therefore a pair of vectors does not def\/ine a solution
to WDVV equations (see also Remark~\ref{remark1} above). The following
propositions deal with the next simplest cases when $A$ consists of~3,~4 and~5
vectors respectively. In fact all irreducible trigonometric $\vee$-systems with up to 5 covectors have to be two-dimensional.

\begin{proposition}\label{proposition2}
Let system $A$ consist of three vectors~$\a$,~$\b$,~$\g$ with nonzero
multiplicities~$c_\a$, $c_\b$,~$c_\g$. Then $A$ is an irreducible
trigonometric $\vee$-system iff $\a \pm \b \pm \g=0$ for some choice
of signs. The non-degeneracy condition for the form \eqref{G} is then
given by $c_\a c_\b + c_\a c_\g + c_\b c_\g\ne 0$.

Any such system $A$ defines the solution \eqref{F} of the WDVV
equations with $\lambda=2(c_\a c_\b + c_\a c_\g + c_\b c_\g)(c_\a
c_\b c_\g)^{-1/2}$.
\end{proposition}

\begin{proof} It follows from relations \eqref{V1} that $\g=\a+\b$ up
to multiplication of some of the vectors by~$-1$. We take a basis
$e^1=\a$, $e^2=\b$ in $\C^2$. The bilinear form \eqref{G} takes the
form $G=c_\a x_1^2+c_\b x_2^2 + c_\g (x_1+x_2)^2$. This form is
non-degenerate if\/f $c_\a c_\b + c_\a c_\g + c_\b c_\g \ne 0$. One
can check that
\[
{e^1}^\vee=\frac{(c_\b+c_\g)e_1-{c_\g}e_2}{c_\a c_\b + c_\a c_\g +
c_\b c_\g}, \qquad {e^2}^\vee=\frac{-c_\g e_1+(c_\a+c_\g)e_2}{c_\a
c_\b + c_\a c_\g + c_\b c_\g},
\]
where $e_1$, $e_2$ is dual basis to $e^1$, $e^2$, that is
$e^i(e_j)=\delta^i_j$.
 Relations \eqref{V1} look as follows
\begin{gather*}
\big({e^1}^\vee,{e^2}^\vee\big)(c_\b+c_\g)+\big({e^1}^\vee,{e^1}^\vee\big)c_\g=0, \\
\big({e^1}^\vee,{e^2}^\vee\big)(c_\a+c_\g)+\big({e^2}^\vee,{e^2}^\vee\big)c_\g=0, \\
\big({e^1}^\vee,{e^2}^\vee\big)(c_\a-c_\b)+\big({e^1}^\vee,{e^1}^\vee\big)c_\a-\big({e^2}^\vee,{e^2}^\vee\big)c_\b=0,
\end{gather*}
and they are automatically satisf\/ied. Relation \eqref{V2} results to
one scalar equation
\[
\frac{\lambda^2}{4}\left(c_\a c_\b (\a^\vee,\b^\vee)+c_\a c_\g
(\a^\vee,\g^\vee)+c_\b c_\g (\b^\vee,\g^\vee)\right)=c_\a c_\b+c_\a
c_\g +c_\b c_\g,
\]
 which has solution as stated in the formulation.
\end{proof}

In the following proposition we study conf\/igurations consisting of
four covectors.

\begin{proposition}\label{proposition3}
Let system $A$ consist of four vectors $\a$, $\b$, $\g$, $\delta$ with
nonzero multiplicities $c_\a$, $c_\b$, $c_\g$, $c_\delta$. Then $A$ is an
irreducible trigonometric $\vee$-system iff the vectors in $A_+$
take the form $e^1$, $e^2$, $e^1 \pm e^2$ in a suitable basis, and the
corresponding multiplicities $c_1$, $c_2$, $c_\pm$ satisfy $c_1=c_2$.
 This property is equivalent to the
orthogonality $(e^1+e^2, e^1-e^2)=0$ under the corresponding
$\vee$-product.
 The non-degeneracy
condition for the form \eqref{G} is then given by
$\Delta=(c_1+2 c_+)(c_1+2c_-)\ne 0$.

These systems $A$ define the solutions \eqref{F} of the WDVV
equations with $\lambda=2\Delta
c_1^{-1/2}(4c_+c_-+c_1(c_++c_-))^{-1/2}$ once $\lambda$ is finite.
\end{proposition}

\begin{proof} It follows from the series relations \eqref{V1} that
there is a a vector $\a \in A$ such that all the remaining vectors
$\b,\g, \delta \in A$ belong to single $\a$-series $\Gamma^1_\a$.

Indeed, otherwise, up to renaming the covectors and taking opposite,
we have $\delta=\gamma+ n \a$, $n \in \N$, $(\a,\b)=(\g,\delta)=0$.
Then consideration of $\b$-series gives $2\g+n\a=m \b$ for some $m
\in \Z$. And consideration of $\gamma$-series gives $\a + p \g = \pm
\b$ for some $p \in \Z$. Therefore $2 \g + n \a = \pm m(\a + p \g)$,
hence $n=\pm m$ and $2 = \pm mp$, thus $m=\pm 1$ or $p=\pm 1$. In
the case $m=\pm 1$ we have $n=1$ hence $\gamma$-series contains
$\delta$ together with $\a$ and $\b$. And in the case $p=\pm 1$ the
$\a$-series contains~$\b$ together with $\g$ and $\delta$.

So we can assume that there is only one $\a$-series so that the
remaining vectors take the form $\gamma=\b + n_1 \a$, $\delta= \b +
n_2 \a$ with integer $n_2 > n_1 >0$. By considering  $\b$-series we
conclude that $n_1=1$. Consider now the $\delta$-series. It is easy
to see that covector $\b$ has to form a single series, therefore
$(\b,\delta)=0$ and the  covectors $\b+\a$ and $\a$ belong to a
$\delta$-series. This is possible only if $n_2=2$. Taking now the
basis vectors as $e^1=\a$, $e^2=\b + \a$ we conclude that the system
$A$ consists of covectors $e^1$, $e^2$, $e^1 \pm e^2$.

The bilinear form \eqref{G} takes now the form
\begin{gather*}
G=c_1 x_1^2+c_2 x_2^2 + c_+ (x_1+x_2)^2+ c_- (x_1-x_2)^2\\
\phantom{G}{}=(c_1+c_++c_-) x_1^2 + (c_2+c_+ + c_-) x_2^2 + 2 (c_+-c_-) x_1 x_2,
\end{gather*}
which has determinant $\Delta= c_1 c_2 +(c_1+c_2)(c_++c_-)+4c_+c_-$.
Therefore \begin{gather} \big(e^1,e^1\big)=\Delta^{-1}(c_2+c_++c_-),\qquad
\big(e^2,e^2\big)=\Delta^{-1}(c_1+c_++c_-),\nonumber\\
\big(e^1,e^2\big)=\Delta^{-1}(c_--c_+).\label{scpr}
\end{gather}

Now we analyze the series relations \eqref{V1}. The orthogonality
$(e^1-e^2, e^1+e^2)=0$ is clearly equivalent to the condition
$c_1=c_2$. Then the remaining conditions \eqref{V1} on $(e^1\pm
e^2)$-series are automatically satisf\/ied. The condition \eqref{V1}
for the $e^1$-series has the form
\[
c_-\big(-e^1+e^2,e^1\big)+ c_2 \big(e^2, e^1\big)+ c_+ \big(e^1+e^2,e^1\big)=0,
\]
and it follows from the scalar products \eqref{scpr}. The condition
on the $e^2$-series is also satisf\/ied.

It is easy to check that relation \eqref{V2} holds if\/f $\lambda$ is as stated, hence proposition is proven.
\end{proof}

\begin{proposition}\label{proposition4}
Let irreducible trigonometric $\vee$-system $A$ consist of five vectors with non-zero multiplicities. Then in the appropriate basis $A_+$ takes the form $e^1$, $2e^1$, $e^2$, $e^1 \pm e^2$ and the corresponding multiplicities $c_1$, $\tilde c_1$, $c_2$, $c_{\pm}$ satisfy $c_+=c_-$ (equivalently, $(e^1,e^2)=0$) and $2\tilde c_1 c_2 = c_+(c_1-c_2)$.

The form \eqref{G} is then non-degenerate when $\Delta=(c_1+4\tilde c_1 + 2 c_+)(c_2+2c_+)\ne 0$. The corresponding solution of the WDVV equations has the form \eqref{F} with $\lambda=\sqrt{2}\Delta (c_2+2c_+)^{-1/2}(c_1+4\tilde c_1)^{-1/2} c_+^{-1/2}$.
\end{proposition}

Proof is obtained by simple analysis of the series conditions \eqref{V1}. One can f\/irstly establish that $A$ is two-dimensional. Then it is easy to see that $A$ has to contain collinear vectors, and the required form follows.

To conclude this section we present a few examples of trigonometric $\vee$-systems on the plane with higher number of vectors. Recall f\/irstly that the positive roots of the root system ${\cal G}_2$ can be written as $\a$, $\b$, $\b+\a$, $\b+n \a$, $\b+(n+1)\a$, $2\b+(n+1)\a$ where $n=2$. One can show that for integer $n>2$ the above vectors never form a trigonometric $\vee$-system, and that for $n=2$ the multiplicities have to satisfy $c_\a=c_{\b+\a}=c_{\b+n\a}$ and $c_\b=c_{2\b+(n+1)\a}=c_{\b+(n+1)\a}$ which is the case of the ${\cal G}_2$ system.

There are though some possibilities to extend the ${\cal G}_2$ system. Firstly, one can show that ${\cal G}_2 \cup {\cal A}_2$ where the system ${\cal A}_2$ consists of doubled short roots of ${\cal G}_2$, is a trigonometric $\vee$-system for appropriate multiplicities. Secondly the following proposition takes place.

\begin{proposition}\label{proposition5} Let $A$ consist of the vectors $e_1$, $e_2$, $2e_2$, $\frac12 (e_1 \pm e_2)$, $\frac12 (e_1\pm 3 e_2)$ with the corresponding nonzero multiplicities $c_1$, $c_2$, $\tilde c_2$, $a$, $b$. Then $A$ is a trigonometric $\vee$-system iff the multiplicities satisfy the relations $a=3b$, $c_2=a+ 2 \tilde c_2$, $(2c_1+b)c_2=(c_1+2b)a$.
\end{proposition}
Note that in the limiting case $\tilde c_2=0$ we recover the system ${\cal G}_2$ with special multiplicities.

An example of trigonometric $\vee$-system with yet higher number of vectors is given by vectors $e_1$, $2e_1$, $e_2$, $2e_2$, $e_1 \pm e_2$, $e_1 \pm 2e_2$, $2e_1 \pm e_2$ where the multiplicities can be chosen appropriately.

\section[Relations with generalized Calogero-Moser-Sutherland systems]{Relations with generalized Calogero--Moser--Sutherland\\ systems}

Relation between $\vee$-systems and the property of a Schr\"odinger
operator of CMS type to have a~factorized eigenfunction was observed
by Veselov in \cite{V3}. Namely, it was shown in \cite{V3} that if
an operator
\[
L=-\Delta+\sum_{\a\in A_+}
\frac{m_{\a}(m_{\a}+1)(\a,\a)}{\sin^2 (\a,x)}
\] has a formal
eigenfunction
\[ \psi=\prod_{\a \in A_+} \sin^{-m_{\a}}(\a,x), \qquad
L \psi =\mu \psi,
\] then $F=\sum\limits_{\a\in A_+} m_\a(\a,x)^2 \log
(\a,x)$ satisf\/ies the WDVV equations. The following theorem
establishes the converse statement in the case of trigonometric
$\vee$-systems.

\begin{theorem} \label{theorem2}
Let $A$ be a trigonometric $\vee$-system consisting
of pairwise non-collinear covectors $\a$ with multiplicities $c_\a$.
Then Schr\"odinger operator \begin{gather}\label{sch}L=-\Delta+\sum_{\a\in A}
\frac{c_{\a}(c_{\a}+1)(\a,\a)}{\sin^2 \a(x)}\end{gather} constructed by the
metric \eqref{G}
 has the  formal
eigenfunction \begin{gather}\label{eig} \psi=\prod_{\a \in A} \sin^{-c_{\a}}\a(x),
\qquad L \psi =\mu \psi. \end{gather}
\end{theorem}

\begin{proof} The property $L\psi = \mu \psi$ is equivalent to the
identity \begin{gather}\label{iden} \sum_{\a \ne \b} c_\a c_\b (\a,\b) \cot \a(x)
\cot \b(x)={\rm const}. \end{gather} To establish the last identity it is
suf\/f\/icient to show that the left-hand side of \eqref{iden} is
non-singular. In other words, we need to show that \begin{gather}\label{trt}
\sum_{\b, \b \ne \a} c_\b (\a,\b) \cot \b(x)=0 \end{gather} if $\cot
\a(x)=0$. The last properties are suf\/f\/icient to check when summation
is taken along arbitrary $\a$-series, then it is guaranteed by
relation \eqref{V1}. This proves the theorem.
\end{proof}

\begin{corollary} \label{corollary1}
Assume that function \eqref{F} constructed by a
set of pairwise non-collinear covectors~$\a$ with multiplicities
$c_\a$ satisfies the WDVV equations~\eqref{wdvv}. Then relation
\eqref{eig} holds for the Schr\"odinger operator~\eqref{sch}.
\end{corollary}

Conversely, the property of a Schr\"odinger operator to have a
factorized eigenfunction implies that the corresponding vectors
$\sqrt{c_\a}\a$ form a rational $\vee$-system \cite{V3}. This
property is also suf\/f\/icient to obtain the trigonometric
$\vee$-system, and the arguments are close to~\cite{V3}.

\begin{theorem}\label{theorem3}
Assume that the Schr\"odinger operator \eqref{sch} has an
eigenfunction~\eqref{eig}.  Then the set $A$ of vectors $\a$ with the
multiplicities $c_\a$ forms the trigonometric $\vee$-system.
\end{theorem}

\begin{proof} From equation \eqref{eig}, \eqref{sch} it follows
identity \eqref{trt} at $\cot \a(x)=0$. Therefore for each
$\a$-series $\Gamma_\a^s$ we have \begin{gather}\label{sersum} \sum_{\b \in \Gamma_\a^s}
c_\b (\a,\b) \a\wedge \b=0. \end{gather} Let $\b^\text{w}$ denote a vector
dual to $\b$ with respect to the inner product $(\cdot,\cdot)$
involved in the Schr\"odinger equation. By summing identities
\eqref{sersum} along all the $\a$-series we conclude that \begin{gather}\label{rav}
\sum_{\b \in A} c_\b \b(\a^\text{w}) \b^\text{w} \quad \mbox{is
proportional to } \a^\text{w}. \end{gather} Now we can decompose the space
$V=V_1\oplus\cdots\oplus V_k$ so that the operator $\sum_{\b\in A}
c_\b \b \otimes \b^\text{w}$ is equal to constant $\mu_i$ on $V_i$. We
can also assume that $(V_i, V_j)=0$ if $i\ne j$. It follows from
\eqref{rav} that $G(\cdot,\cdot)|_{V_i}=\mu_i (\cdot,\cdot)|_{V_i}$.
Therefore identities \eqref{sersum} imply
\[ \sum_{\b \in \Gamma_\a^s}
c_\b \a(\b^\vee) \a\wedge \b=0
\]
which are identities \eqref{V1} from the def\/inition of the
trigonometric $\vee$-systems.
\end{proof}

\begin{corollary}\label{corollary2}  Assume  that the Schr\"odinger operator
\eqref{sch} has an eigenfunction \eqref{eig}. Assume also that the
system $A$ is irreducible and that for some $\Lambda$ and any
$a,b,c,d \in V$ the property \begin{gather}\label{fin}
 \sum_{\a,\b \in A_+}
(\Lambda  (\a, \b) - 1)c_\a c_\b  B_{\a,\b}(a,b)
B_{\a,\b}(c,d)  =0 \end{gather} holds. Then the corresponding function
\eqref{F} with appropriate $\lambda$ satisfies the WDVV equations~\eqref{wdvv}.
\end{corollary}

\begin{remark}\label{remqrk4} The previous corollary also holds for the reducible
systems $A$ if we replace the Schr\"odinger equation metric
$(\a,\b)$ in \eqref{fin} by the $\vee$-product $\a(\b^\vee)$. In this case $\lambda=2 \sqrt{\Lambda}$.
\end{remark}

\section{Concluding remarks}

Trigonometric $\vee$-systems require further investigations. It
would be interesting to obtain almost dual prepotentials for the
Frobenius manifolds of the af\/f\/ine Weyl groups as well as for their
discriminants (cf.\ rational case~\cite{D2,FV1}).
Comparison with a recent work on the elliptic solutions~\cite{Str}
might also be interesting. We also hope that the series conditions
would allow understanding and eventually classif\/ication of the
trigonometric $\vee$-systems. We hope to return to some of these
questions soon.

\subsection*{Acknowledgements}

I am very grateful to L.~Hoevenaars, A.~Kirpichnikova, M.~Pavlov, I.~Strachan and A.P.~Veselov for useful
and stimulating  discussions. The work was partially supported by
the EPSRC grant EP/F032889/1, by European research network ENIGMA
(contract MRTN-CT-2004-5652), by PMI2 Project funded by the UK
Department for Innovation, Universities and Skills for the benef\/it
of the Japanese Higher Education Sector and the UK Higher Education
Sector.

\pdfbookmark[1]{References}{ref}
\LastPageEnding

\end{document}